\documentclass[aps,prl,preprint,showpacs,superscriptaddress]{revtex4-1}

\usepackage{graphicx}  
\usepackage{dcolumn}   
\usepackage{bm}        
\usepackage{amssymb}   
\usepackage{amsmath}
\usepackage{esint}
\usepackage{hyperref}

\hypersetup{%
   colorlinks=true,
   urlcolor=blue,
   linkcolor=blue,
   citecolor= blue,
   raiselinks=false
}

\begin{document}


\title{Measurement of mechanical deformations induced by enhanced electromagnetic stress on a parallel metallic-plate system}

\affiliation{Department of Physics, The Hong Kong University of Science and Technology, Clear Water Bay, Kowloon, Hong Kong, China}
\affiliation{William Mong Institute of Nano Science and Technology, The Hong Kong University of Science and Technology, Clear Water Bay, Kowloon, Hong Kong, China}
\affiliation{Department of Physics, City University of Hong Kong, Tat Chee Avenue, Kowloon, Hong Kong, China}

\author{M. Wang} \affiliation{Department of Physics, The Hong Kong University of Science and Technology, Clear Water Bay, Kowloon, Hong Kong, China} \affiliation{William Mong Institute of Nano Science and Technology, The Hong Kong University of Science and Technology, Clear Water Bay, Kowloon, Hong Kong, China}

\author{S. Wang} \affiliation{Department of Physics, The Hong Kong University of Science and Technology, Clear Water Bay, Kowloon, Hong Kong, China} \affiliation{Department of Physics, City University of Hong Kong, Tat Chee Avenue, Kowloon, Hong Kong, China}

\author{Q. Zhang} \affiliation{Department of Physics, The Hong Kong University of Science and Technology, Clear Water Bay, Kowloon, Hong Kong, China} \affiliation{William Mong Institute of Nano Science and Technology, The Hong Kong University of Science and Technology, Clear Water Bay, Kowloon, Hong Kong, China}

\author{C. T. Chan} \affiliation{Department of Physics, The Hong Kong University of Science and Technology, Clear Water Bay, Kowloon, Hong Kong, China} 

\author{H. B. Chan} \email {hochan@ust.hk} \affiliation{Department of Physics, The Hong Kong University of Science and Technology, Clear Water Bay, Kowloon, Hong Kong, China} \affiliation{William Mong Institute of Nano Science and Technology, The Hong Kong University of Science and Technology, Clear Water Bay, Kowloon, Hong Kong, China}

\date{\today}

\begin{abstract}
We measured the electromagnetic stress-induced local strain distribution on a centimeter-sized parallel-plate metallic resonant unit illuminated with microwave. Using a fiber interferometer, we found that the strain changes sign across the resonant unit, in agreement with theoretical predictions that the attractive electric and repulsive magnetic forces act at different locations. The enhancement of the corresponding maximum local electromagnetic stress is stronger than the enhancement of the net force, reaching a factor of \textgreater 600 compared to the ordinary radiation pressure.
\end{abstract}

\maketitle


Recent advances in metamaterials have opened up a new paradigm for manipulating light or sound using functionalities not achievable with conventional materials. Familiar examples include negative refraction \cite{ref1, ref2} and cloaking \cite{ref2, ref3}. Active tunability of electromagnetic (EM) properties of metamaterials holds promise in taking the wave manipulation functionality to the next level. Such active control can be realized by manipulating the shape or the relative positions of the building blocks \cite{ref4, ref5}. Various actuation mechanisms that utilize thermal \cite{ref6, ref7}, electrostatic \cite{ref8, ref9, ref10}, magnetic \cite{ref11} or mechanical \cite{ref12} effects have been proposed or demonstrated. Alternatively, one can also exploit the mechanical effects induced by the EM radiation \cite{ref13,ref14,ref15,ref16,ref17,ref18,ref19,ref20,ref21,ref22,ref23,ref24}. For most frequencies, the radiation pressure exerted by the EM field on metamaterials is small. However, at resonance the EM field can be strongly concentrated at certain locations of the metamaterial elements. The associated mechanical effects could be significantly enhanced, opening new opportunities for nonlinear \cite{ref18, ref25} or reconfigurable \cite{ref6, ref7, ref8, ref9, ref11} metamaterials.

While it is well-established that the net EM force acting on a resonating element by a time-harmonic external field can be much larger than the ordinary radiation force \cite{ref13, ref15, ref16, ref22, ref24, ref26, ref27}, there are recent predictions that enhancements in the local EM stress can be even stronger \cite{ref28}. Such notions are based on the fact that the electric and magnetic forces, which are generated by the oscillating charges and currents, respectively, tend to have opposite directions at resonance. Since these two forces act on different parts of the resonating element, the local stress is expected to significantly exceed the average pressure which is in fact a remnant of the imperfect cancellation of the electric and magnetic forces \cite{ref15}. 
	
Apart from the practical goal of generating a large mechanical response, measurement of the local EM stress exerted on metamaterials is also of fundamental interest. It is well-known that the total time-averaged EM force on an isolated object can be calculated by integrating the EM stress tensor across any boundary that completely encloses the object. However, theoretical analysis has shown that for certain metamaterials, the calculation of the force density inside the metamaterial is not straight forward but requires additional information (such as electrostrictive tensor components) that goes beyond standard effective constitutive parameters \cite{ref29, ref30}. A rigorous approach of calculating the EM local stress using macroscopic field does not yet exist. In some cases, such difficulty can be circumvented if the EM field is negligible inside the material \cite{ref28}.  Measurement of the EM stress, on the other hand, is often complicated by thermal effects that can dominate the mechanical response \cite{ref22, ref24}. To our knowledge, the local EM stress exerted on a resonating unit of a metamaterial has not yet been measured.  
	
In this Letter, we report measurement of the deformation of a system consisting of centimeter-sized gold parallel plates induced by the time-averaged EM stress of incident microwave. The local deformation is measured by scanning a fiber interferometer across the top plate. We distinguish contributions of the EM stress induced by the microwave from thermal effects that generate an additional phase lag in the mechanical vibrations \cite{ref31}. The measured EM strain distribution is found to change sign across the plate, in agreement with theoretical predictions \cite{ref28} that the attractive electric force and the repulsive magnetic force are concentrated at different locations. At the microwave resonance frequency, the net EM force on the plate is enhanced by a factor of $\sim$20 relative to the conventional radiation force while the enhancement of EM stress reaches a factor of \textgreater 600.

\begin{figure}
\includegraphics[scale=1.5]{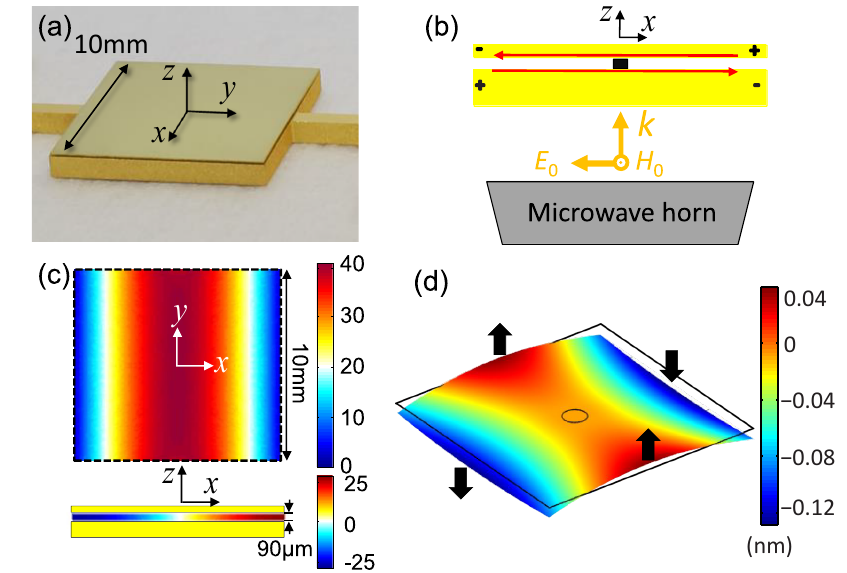}
\caption{\label{fig:1}  (color online). (a) The double-plate resonating unit. The thick bottom plate is connected to the support by two rods. (b) Cross-sectional view of the double plate system (not to scale), showing induced anti-parallel currents (red arrows) and charges at resonance. A silicon post supports the top plate at the middle. (c) The magnetic field (along the $y$ direction) distribution in the gap at resonance (top) and the corresponding electric field (along the $z$ direction) distribution along the middle of the plate (bottom). Both are normalized to the incident field amplitude. (d) Calculated deformation of the top plate in response to the time-averaged electromagnetic stress. The black arrows represent the directions of the stress at the middle of the four edges.}
\end{figure}

Figure \hyperref[fig:1]{1(a)} shows the parallel-plate resonant unit in our experiment. It consists of a thin plate of thickness 68 $\mu$m made entirely of gold and a copper bottom plate that is much thicker (1 mm) and uniformly coated with 1.2 $\mu$m gold. Both plates measure 10 mm by 10 mm in the $x$-$y$ plane. As assembled, the gap between the two plates is $\sim$90 $\mu$m, defined by a cylindrical silicon column of radius 500 $\mu$m that is glued to the two plates. The two plates serve as the resonant cavity that confines the EM field. Only the thin top plate shows detectable deformation. Two rods are connected to the left and right edges of the thick bottom plate [Fig. \hyperref[fig:1]{1(a)}], providing mechanical support as well as paths for conducting away the heat generated when the microwave radiation is turned on, as we will describe later.

As shown in the bottom part of Fig. \hyperref[fig:1]{1(b)}, a microwave horn is placed 5 mm below the bottom plate. It emits microwave with electric field, magnetic field and wave vector along the $x$, $y$ and $z$ directions, respectively. At resonance, the EM energy stored between the two plates attains a maximum. Figure \hyperref[fig:1]{1(c)} shows the calculated magnetic field and electric field in the gap for the antisymmetric mode of the cavity where anti-parallel oscillating currents are generated on the two plates due to the time-varying magnetic flux in the $y$ direction. Because of the current flow, opposite charges accumulate periodically on the edges of the plates, as shown in Fig. \hyperref[fig:1]{1(b)}. The anti-parallel currents generate a repulsive magnetic force near the middle of the plates while the opposite accumulated charges lead to an attractive electric force near the edges. Previous works focused on the enhancement of the net force that results from the incomplete cancellation of the time-averaged attractive electric and repulsive magnetic forces \cite{ref22}. Our goal here is to measure the local stress, and to demonstrate that local forces can achieve even stronger enhancement by exploiting the different spatial distribution of the electric and magnetic forces. Calculations (Supplementary Information) indicate that the maximum stress is enhanced by a factor of more than 600 times over the ordinary photon pressure, compared to the enhancement factor of about 20 for the net EM force. Figure \hyperref[fig:1]{1(d)} shows the calculated deformation of the plate in response to the stress exerted by the EM field. With the center of the plate fixed by the supporting post, the left and right edges bend downwards due to the electric force while the top and bottom edges bend upwards due to the magnetic force.

\begin{figure}
\includegraphics[scale=1.5]{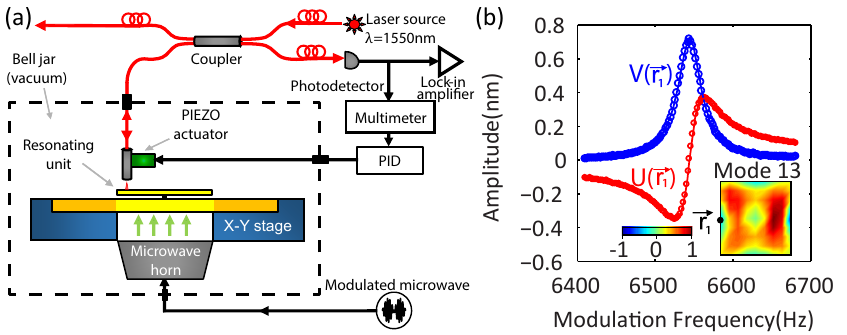}
\caption{\label{fig:2} (color online). (a) Local deformation of the top plate is measured by a fiber interferometer. The location of the parallel-plate unit relative to the optical fiber is changed by an X-Y positioning stage. (b) Vibration amplitude of the top plate in-phase ($U$, red) and out-of-phase ($V$, blue) with the intensity modulations of the microwave at location $\vec r_1$ for normal mode 13. Inset: the measured mode shape for mode 13, normalized to the maximum vibration amplitude across the plate.}
\end{figure}

A fiber interferometer working at wavelength of 1550 nm is used to measure the deformation at different locations across the plate (Supplementary Information). As shown in Fig. \hyperref[fig:2]{2(a)}, the device and the microwave horn are scanned along the X-Y direction by positioners. By maximizing the mechanical response, the microwave resonance frequency of the parallel-plate system is determined to be 14.57 GHz. The intensity of the microwave radiation is modulated at $\omega$, generating a periodic stress on the top plate. Vibrations of the top plate leads to periodic modulations of the reflected light intensity in the fiber interferometer. The light intensity is measured with a photodetector, the output of which is connected to a lockin amplifier referenced at $\omega$. Measurements are performed at room temperature and pressure of $<$ 10$^{-5}$ torr.

The equation of motion for the top plate is \cite{ref32}:

\begin{eqnarray}
D\nabla^4A+\rho h\frac{\partial^{2}A}{\partial t^{2}}+\gamma \frac{\partial A}{\partial t}=P(\vec r,t)
\label{eq:1}.
\end{eqnarray}

\noindent where $A(\vec{r},t)$ is the local displacement perpendicular to the substrate, $\rho$ is the mass density, $h$ is the thickness of the plate and $\gamma$ characterizes the damping. $D=Y_gh^3/[12(1-\nu^2)]$ is the flexural rigidity, with $Y_g$ being the Young$'$s modulus and $\nu$ being the Poisson ratio. Since the radius of the silicon stub supporting the plate is much smaller than the plate width, we consider free vibrations of the plate with a nodal point at the plate center. For periodic excitation $P(\vec r,t)=P_0(\vec r)\cos(\omega t)$ at a modulation frequency $\omega$, $A(\vec r,t)$ is given by:

\begin{eqnarray}
A(\vec{r},t)=U(\vec{r})\cos(\omega t)-V(\vec r)\sin(\omega t)
\label{eq:2}.
\end{eqnarray}

\noindent where $U(\vec{r})$ and $V(\vec{r})$ are the spatial distribution of the amplitude of vibrations in-phase and out-of-phase with $P$. The response of the plate can be written as a linear combination of all the normal modes:

\begin{eqnarray}
U(\vec{r},\omega)+iV(\vec{r},\omega)= \sum\nolimits_n  \frac{A_n(\vec r)}{1-(\omega/\omega_n)^2+i(^\omega/_{\omega_nQ_n})}%
\label{eq:3}.
\end{eqnarray}

\noindent where $\omega_n$ and $Q_n=\rho h\omega_n/\gamma$ are the resonance frequency and the effective quality factor of the $n^{th}$ mode, respectively. $A_n(\vec r)$ represents the contributions of the $n^{th}$ mode. In our system, $Q_n$'s are rather large (\textgreater120) so that when $\omega$ is close to $\omega_n$, the response of the plate is dominated by mode $n$. Figure \hyperref[fig:2]{2(b)} plots $U(\vec r_1)$ and $V(\vec r_1)$ measured at one particular location $\vec r_1$ at the left edge of the plate (marked in the inset) for mode 13. Since the displacement of mode 13 is negative (towards the thick plate) at $\vec r_1$, both $U(\vec r_1)$ and $V(\vec r_1)$ pick up an extra negative sign when compared to the response of a driven harmonic oscillator.

\begin{figure}
\includegraphics[scale=1.2]{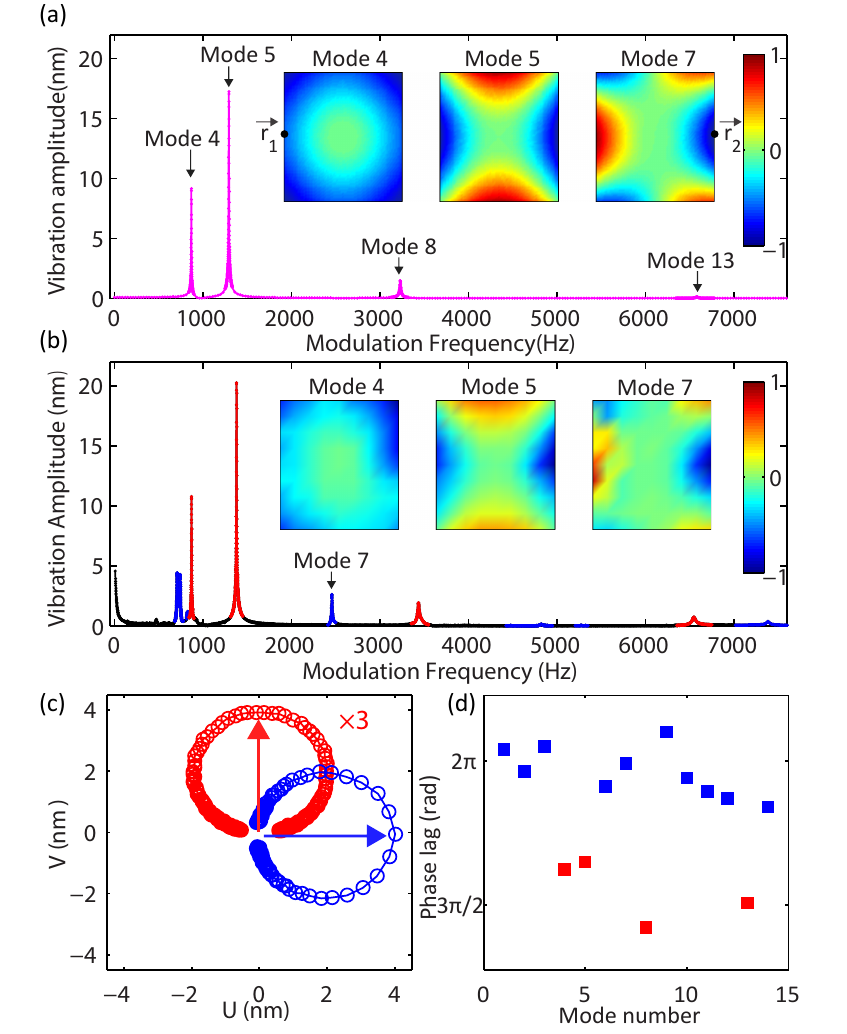}
\caption{\label{fig:3} (color online). (a) Calculated vibration amplitude at location $\vec r_1$ (marked by the black dot in the insets) in response to the electromagnetic stress as a function of microwave modulation frequency. Insets: the mode profiles for normal modes 4, 5 and 7 for the 8 mm by 8 mm area at the center of the plate, normalized to the maximum vibration amplitude across the plate. (b) Measured vibration amplitude. The red peaks represent modes with the same symmetry as the electromagnetic stress. The blue peaks represent modes with the opposite symmetry that are missing in the main panel in (a). Insets: measured mode profiles for modes 4, 5, and 7. (c) Response of modes 13 (red) and 7 (blue) at location $\vec r_2$ plotted in the U-V phase space. The red (blue) arrow indicates the vibrations lag the microwave modulation by $3\pi/2(2\pi)$. (d) Measured phase lags of the first 14 normal modes, plotted in red/blue for modes with the same/different symmetry as the EM stress. Error bars are comparable to the symbol size.}
\end{figure}

The insets of Fig. \hyperref[fig:3]{3(a)} shows the normalized spatial distribution $s_n(\vec r)$ of a few modes ($n$ = 4, 5, 7) calculated with finite element analysis. They are in good agreement with the measured $s_n(\vec r)$ that is excited by the incident microwave with intensity modulated at $\omega_n$ [insets in Fig. \hyperref[fig:3]{3(b)}]. Unless otherwise stated, the microwave frequency is fixed at the resonant value of 14.57 GHz. The microwave power is 250 mW.

The main plot in Fig. \hyperref[fig:3]{3(a)} shows the $\omega$ dependence of the calculated vibration amplitude $R=\sqrt{U^2(\vec r_1,\omega)+V^2(\vec r_1,\omega)}$ at a point $\vec r_1$ on the edge of the plate (marked by the black dot in the inset). For each mode, $A_n(\vec r)$  is determined by the overlap of the mode profile $s_n(\vec r)$ with $P_0(\vec r)$:

\begin{eqnarray}
A_n(\vec r)=\frac{1}{\rho h\omega^2_n}s_n(\vec r) \iint s_n(\vec {r'})P_0(\vec {r'})dx'dy'
\label{eq:4}.
\end{eqnarray}

\noindent In other words, the relative heights of the peaks in Fig. \hyperref[fig:3]{3(a)} depend on the spatial distribution of the applied stress $P_0(\vec r)$. Since the stress exerted by the microwave radiation is symmetric about both the X and Y axes, only vibration modes with the same symmetry can be excited. Figure \hyperref[fig:3]{3(b)} shows the measurements, with these symmetric modes plotted in red. The anti-symmetric modes (anti-symmetric about the X and/or Y axis) are plotted in blue. They are dark modes in the numerical simulations, as shown by their absence in Fig. \hyperref[fig:3]{3(a)}. In experiments, our device is not perfectly symmetric as neither the gold plate thickness nor the gap is exactly uniform. Therefore, the anti-symmetric modes can also be excited, as plotted in blue in Fig. \hyperref[fig:3]{3(b)}.

One approach to obtain the strain induced by the EM wave is to measure the response at dc (i.e. zero modulation frequency). However, we find that as $\omega$ is reduced towards zero, photothermal effects become dominant, as evident by the rise at low frequencies depicted in black in Fig. \hyperref[fig:3]{3(b)}. At the microwave resonance, the radiation exerts strong EM stress $P_{EM}$ that originates from the enhanced current and charge oscillations. This current also leads to ohmic heating and thermal expansion. The thermal deformations induced by radiation are commonly associated with photothermal forces. Unlike the EM force that appears instantaneously once the radiation is turned on, the mechanical deformation induced by the photothermal force exhibits a delay. Assuming a delayed impulse response of $h(t)=1-$exp$(-t/\tau)$, the photothermal stress can be written as \cite{ref31}:

\begin{eqnarray}
P_{thermal}(\vec r, t)=\int_0^t {c(\vec r,t)\frac{dI(t')}{dt'}h(t-t')dt'}
\label{eq:5}.
\end{eqnarray}

\noindent where $I$ is the intensity of the microwave and $c$ characterizes the spatial distribution of the photothermal stress. With the contribution of the photothermal stress included, Eq. \hyperref[eq:3]{(3)} is modified to:

\begin{eqnarray}
U(\vec{r},\omega)+iV(\vec{r},\omega)= \sum\nolimits_n  \frac{A_{EM,n}(\vec r)+\frac{A_{thermal,n}(\vec r)}{(1+i\omega\tau)}}{1-(\omega/\omega_n)^2+i(^\omega/_{\omega_nQ_n})}%
\label{eq:6}.
\end{eqnarray}

\noindent where $A_{EM,n}$ and $A_{thermal,n}$ characterize the vibration amplitude of the $n^{th}$ mode excited by the EM stress and photothermal stress, respectively. The sharp increase of the vibration amplitude as $\omega$ is lowered towards zero indicates that in our device the strain induced by the photothermal stress is much larger than that of the EM stress at dc. By fitting to the low frequency region in Fig. \hyperref[fig:3]{3(b)}, we find that $\tau$ = 132 ms. Measurements at other locations yield similar values.

Our device is designed to have mode eigenfrequencies much higher than $1/\tau$ (e.g. $\omega_1$ = 708 Hz) so that $\omega_n\tau\gg1$ for all modes. At modulation frequency $\omega$$\sim$$\omega_n$, the photothermal term in the numerator of Eq. \hyperref[eq:6]{(6)} is reduced to $-iA_{thermal,n}(\vec r)/(\omega\tau)$. Apart from decreasing rapidly with $\omega$, the photothermal contribution to the vibrations lags behind that induced directly by the EM stress by an extra phase of $\pi/2$. Figure \hyperref[fig:3]{3(c)} compares the phase of vibrations of modes 13 and 7 at location $\vec r_2$ where the displacement is negative (towards the thick plate). For mode 13, the phase of $3\pi/2$ is consistent with the notion that vibrations are excited largely by the EM stress because this mode possesses the same symmetry [inset of Fig. \hyperref[fig:2]{2(b)}] as the EM stress. In contrary, mode 7 has odd symmetry about the $y$ axis [right inset in Fig. \hyperref[fig:3]{3(a)}]. Interestingly, the phase of vibrations lags behind mode 13 by $\pi/2$. Figure \hyperref[fig:3]{3(d)} shows that in general, the phase lag at the eigenfrequencies of the anti-symmetric modes (blue squares) relative to the modulation is close to $2\pi$, larger than that of the symmetric modes by $\sim$$\pi/2$ (red squares). The symmetric and anti-symmetric modes are therefore predominately excited by the EM stress and the photothermal effect, respectively (Supplementary Information).  
 
 The extra $\pi/2$ phase lag of the photothermal response in our system for $\omega>1/\tau$ allows us to exclude the photothermal contribution to the deformation for each mode and extrapolate the remaining part back to zero frequency to obtain the strain induced solely by the EM stress. Specifically, we set the modulation frequency to $\omega_n$ so that the complex amplitude of vibrations described by the series summation in Eq. \hyperref[eq:6]{(6)} are dominated by a single mode $n$:
 
\begin{eqnarray}
  \begin{aligned}
	 U & (\vec r, \omega_n)=-Q_n A_{thermal,n}(\vec r)/(\omega_n\tau) \\
     V & (\vec r, \omega_n)=-Q_n A_{EM,n}(\vec r)          
	\label{eq:7}.
 \end{aligned}
\end{eqnarray}

\noindent Equation \hyperref[eq:7]{(7)} shows that $A_{EM,n}(\vec r)$ can be obtained by dividing the measured vibration amplitude $V(\vec r,\omega_n)$ out of phase with the microwave modulation at $\omega_n$ by the quality factor $Q_n$. By measuring $A_{EM,n}(\vec r)$ for all the modes and setting both $\omega$ and $A_{thermal,n}(\vec r)$ in Eq. \hyperref[eq:6]{(6)} equal to zero, the deformation $A_{EM}(\vec r)$ induced solely by the EM stress can be obtained:

 \begin{eqnarray}
	A_{EM}(\vec r)=\sum\nolimits_n A_{EM,n}(\vec r)       
	\label{eq:8}.
\end{eqnarray}

\begin{figure}
\includegraphics[scale=1.2]{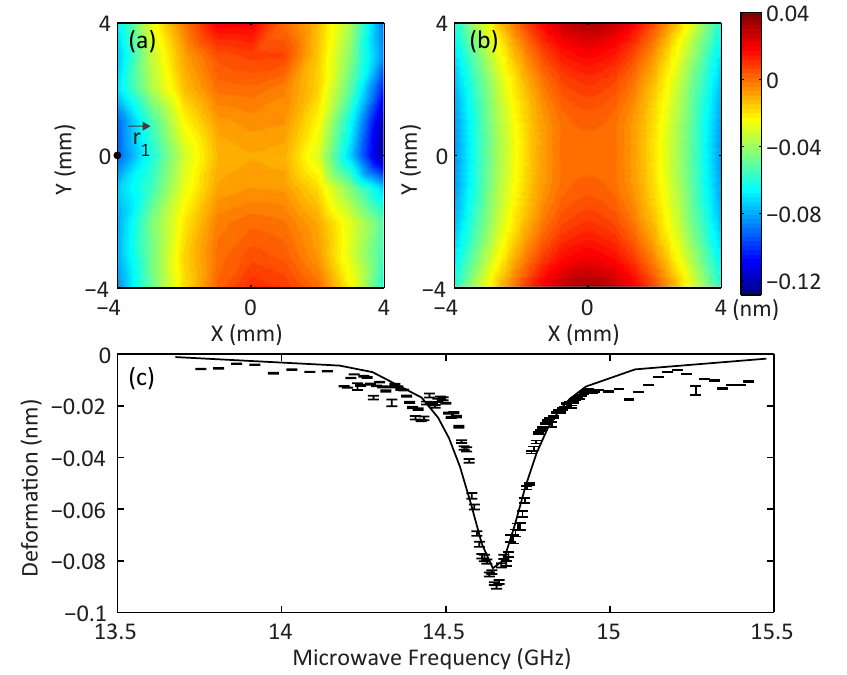}
\caption{\label{fig:4} (color online). (a) Measured and (b) calculated strain of the top plate induced by the electromagnetic stress. (c) Dependence of the deformation of the top plate at position $\vec r_1$ on the microwave frequency. The solid line shows the calculated results.  }
\end{figure}

Figure \hyperref[fig:4]{4(a)} shows $A_{EM}(\vec r)$ measured using the above procedure, with the summation up to $n$ = 14. The largest contribution comes from the symmetric modes, especially modes 4 and 5 (Supplementary Information). The strain distribution agrees well with calculations [Fig. \hyperref[fig:4]{4(b)}] (Supplementary Information). In particular, the sign of the deformation changes for different locations on the plate. The induced currents concentrated near the middle of the upper and lower edges leads to repulsive magnetic forces while charges on the left and right edges generate attractive electric forces.

The time-averaged stress exerted by the EM radiation, in principle, can be deduced by inserting the measured $A_{EM}(\vec r)$ to the equation of motion of the plate [Eq. \hyperref[eq:1]{(1)}] and setting the partial time derivatives to zero. However, this procedure involves taking fourth order derivatives and the calculated stress from our data is too noisy for any meaningful discussion. Instead, we draw a number of conclusions on the EM force/stress based on the good agreement between the measured and predicted strain. First, the imperfect cancellation of the attractive and repulsive stress produces a net EM force of 24.8 nN (for 250 mW incident microwave power), about 20 times larger than ordinary photon force due to simple reflection of photons. Second, our experiment demonstrates that the deformation of the plate is consistent with local EM stress that shows even larger enhancement. The maximum stress, exerted at location $\vec r_1$ in Fig. \hyperref[fig:4]{4(a)}, is more than 600 times larger than ordinary photon pressure. These enhancements only take place at the microwave resonance when the EM waves are concentrated between the two plates. As shown in Fig. \hyperref[fig:4]{4(c)}, as the microwave frequency is tuned away from resonance, the deformation of the plate drastically decreases, in agreement with calculations. The measurements are in good agreement with the calculations in which the width of the square plate is chosen to be 9.946 mm, about 0.5\% smaller than the nominal value

In summary, our experiment demonstrated that the local strain induced by the electromagnetic field on a parallel-plate resonating system is of different sign across the plate. At certain locations, the corresponding EM induced stress is significantly stronger than the already-enhanced average pressure. We showed that EM induced stress can be characterized accurately, and in particular, our work provides a general recipe for isolating the EM stress from the photothermal stress. A better understanding of how light can deform resonating mechanical elements can open new opportunities in tunable and nonlinear meta-materials. Similar resonant enhancement of the EM stress is also expected at optical frequencies for plasmonic cavities \cite{ref16, ref24}. With proper designs, strong optomechanical coupling could be generated. 

This work is supported by Grant No. AoE/P-02/12 from the Research Grants Council of Hong Kong SAR. S. W. is also supported by a grant from City University of Hong Kong (Project No. 9610388).

\section{Supplementary Material}
\subsection{Measurement of local strain by fiber interferometer}
The local deformation of the top plate is measured with a fiber interferometer at wavelength of 1550 nm [Fig. \hyperref[fig:2]{2(a)}]. Light reflected from the top plate and the cleaved end of fiber interfere and leads to intensity modulations that depend on the local plate displacement. The optical fiber is attached to a piezo-electric actuator that maintains the time-averaged separation $d$ between the vibrating plate and the fiber at a constant value through a PID feedback loop as the parallel-plate system is scanned in the X-Y direction by the positioners. The separation is chosen to give the maximum slope in the dependence of the reflected light intensity on $d$. Output of the photodetector is measured by a lockin amplifier that is referenced to the modulation frequency $\omega$ of the microwave. Measurements were performed on an 8 mm by 8 mm square area for two reasons. First, when the fiber approaches the edges of the plate, the distance feedback may damage the fiber. Second, during fabrication a polishing step is necessary to reduce the surface roughness of the gold plate. While the 8mm by 8mm area in the center is clean and has high reflectance, some particles are generated at the edge regions where the reflectance is degraded.

\subsection{Calculation of electromagnetic stress and induced strain}
The calculation of the stress exerted by the microwave radiation includes two steps. First, the EM fields are calculated for exactly the geometry in the experiment using COMSOL. Impedance boundary condition is applied taking into account the finite conductivity of gold at the microwave frequency of 14.57 GHz. The field-induced stress is then calculated using the Maxwell stress tensor approach.

\subsection{Excitation of the anti-symmetric modes by photothermal forces}

The local strain $A_{EM}(\vec r)$ induced by the EM radiation is obtained by summing up the contributions of each mode $A_{EM,n}(\vec r)$ using Eq. \hyperref[eq:8]{(8)}. Figure \hyperref[fig:S1]{S1} plots the contribution of each mode, up to $n$ = 14. The largest contribution comes from the symmetric modes, in particular modes 4 and 5.

\renewcommand{\thefigure}{S1}

\begin{figure} [!htb]
\includegraphics[scale=2]{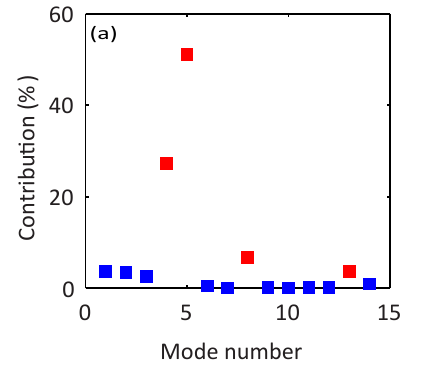}
\caption{\label{fig:S1} The contribution of the normal modes to the electromagnetic strain in Eq. \hyperref[eq:8]{(8)}, plotted in red/blue for modes with the same/different symmetry as the EM stress. Error bars are comparable to the symbol size.  }
\end{figure}

Figure  \hyperref[fig:3]{3(d)} shows that the phase of the anti-symmetric modes lags the symmetric ones by $\pi/2$, indicating that the former are excited largely by the photothermal effects of the EM radiation. One plausible reason for the photothermal contribution being more effective in exciting the anti-symmetric modes is that the thermal conductivity depends on the bulk properties of the metallic elements. For example, the thickness of the gold plate varies by up to 10\% across its area. In contrast, the EM stress originates from currents and charges that are concentrated near the surface and is less susceptible to device non-uniformity.

\subsection{Comparison of the enhancement of the net electromagnetic force to an earlier experiment}
In an earlier experiment \cite{ref22}, the net electromagnetic force on a metal plate with similar lateral size was detected by measuring the capacitance change to a fixed electrode. At microwave resonance, the force was measured to be enhanced by a factor of $\sim$ 100. In this paper, the observed force enhancement is smaller due to changes in the sample configuration. First, the bottom plate in the earlier experiment is a gold-coated silicon wafer that is laterally much larger than the top plate. In contrast, the top and bottom plates in the current experiment have equal lateral size. Second, the direction of propagation of microwave is different. In the first experiment, the microwave horn was placed on top of the double-plate system. For the current experiment, it is necessary to move the horn to the bottom side to give access of the top plate to the fiber interferometer.

\bibliographystyle{apsrev4-1}

\begin{thebibliography}{32}%
\makeatletter
\providecommand \@ifxundefined [1]{%
 \@ifx{#1\undefined}
}%
\providecommand \@ifnum [1]{%
 \ifnum #1\expandafter \@firstoftwo
 \else \expandafter \@secondoftwo
 \fi
}%
\providecommand \@ifx [1]{%
 \ifx #1\expandafter \@firstoftwo
 \else \expandafter \@secondoftwo
 \fi
}%
\providecommand \natexlab [1]{#1}%
\providecommand \enquote  [1]{``#1''}%
\providecommand \bibnamefont  [1]{#1}%
\providecommand \bibfnamefont [1]{#1}%
\providecommand \citenamefont [1]{#1}%
\providecommand \href@noop [0]{\@secondoftwo}%
\providecommand \href [0]{\begingroup \@sanitize@url \@href}%
\providecommand \@href[1]{\@@startlink{#1}\@@href}%
\providecommand \@@href[1]{\endgroup#1\@@endlink}%
\providecommand \@sanitize@url [0]{\catcode `\\12\catcode `\$12\catcode
  `\&12\catcode `\#12\catcode `\^12\catcode `\_12\catcode `\%12\relax}%
\providecommand \@@startlink[1]{}%
\providecommand \@@endlink[0]{}%
\providecommand \url  [0]{\begingroup\@sanitize@url \@url }%
\providecommand \@url [1]{\endgroup\@href {#1}{\urlprefix }}%
\providecommand \urlprefix  [0]{URL }%
\providecommand \Eprint [0]{\href }%
\providecommand \doibase [0]{http://dx.doi.org/}%
\providecommand \selectlanguage [0]{\@gobble}%
\providecommand \bibinfo  [0]{\@secondoftwo}%
\providecommand \bibfield  [0]{\@secondoftwo}%
\providecommand \translation [1]{[#1]}%
\providecommand \BibitemOpen [0]{}%
\providecommand \bibitemStop [0]{}%
\providecommand \bibitemNoStop [0]{.\EOS\space}%
\providecommand \EOS [0]{\spacefactor3000\relax}%
\providecommand \BibitemShut  [1]{\csname bibitem#1\endcsname}%
\let\auto@bib@innerbib\@empty
\bibitem [{\citenamefont {Pendry}(2004)}]{ref1}%
  \BibitemOpen
  \bibfield  {author} {\bibinfo {author} {\bibfnamefont {J.~B.}\ \bibnamefont
  {Pendry}},\ }\href@noop {} {\bibfield  {journal} {\bibinfo  {journal}
  {Contemp. Phys.}\ }\textbf {\bibinfo {volume} {45}},\ \bibinfo {pages} {191}
  (\bibinfo {year} {2004})}\BibitemShut {NoStop}%
\bibitem [{\citenamefont {Shelby}\ \emph {et~al.}(2001)\citenamefont {Shelby},
  \citenamefont {Smith},\ and\ \citenamefont {Schultz}}]{ref2}%
  \BibitemOpen
  \bibfield  {author} {\bibinfo {author} {\bibfnamefont {R.~A.}\ \bibnamefont
  {Shelby}}, \bibinfo {author} {\bibfnamefont {D.~R.}\ \bibnamefont {Smith}}, \
  and\ \bibinfo {author} {\bibfnamefont {S.}~\bibnamefont {Schultz}},\
  }\href@noop {} {\bibfield  {journal} {\bibinfo  {journal} {Science}\ }\textbf
  {\bibinfo {volume} {292}},\ \bibinfo {pages} {77} (\bibinfo {year}
  {2001})}\BibitemShut {NoStop}%
\bibitem [{\citenamefont {Chen}\ \emph {et~al.}(2010)\citenamefont {Chen},
  \citenamefont {Chan},\ and\ \citenamefont {Sheng}}]{ref3}%
  \BibitemOpen
  \bibfield  {author} {\bibinfo {author} {\bibfnamefont {H.}~\bibnamefont
  {Chen}}, \bibinfo {author} {\bibfnamefont {C.~T.}\ \bibnamefont {Chan}}, \
  and\ \bibinfo {author} {\bibfnamefont {P.}~\bibnamefont {Sheng}},\
  }\href@noop {} {\bibfield  {journal} {\bibinfo  {journal} {Nat. Mater.}\
  }\textbf {\bibinfo {volume} {9}},\ \bibinfo {pages} {387} (\bibinfo {year}
  {2010})}\BibitemShut {NoStop}%
\bibitem [{\citenamefont {Zheludev}\ and\ \citenamefont
  {Kivshar}(2012)}]{ref4}%
  \BibitemOpen
  \bibfield  {author} {\bibinfo {author} {\bibfnamefont {N.~I.}\ \bibnamefont
  {Zheludev}}\ and\ \bibinfo {author} {\bibfnamefont {Y.~S.}\ \bibnamefont
  {Kivshar}},\ }\href@noop {} {\bibfield  {journal} {\bibinfo  {journal} {Nat.
  Mater.}\ }\textbf {\bibinfo {volume} {11}},\ \bibinfo {pages} {917} (\bibinfo
  {year} {2012})}\BibitemShut {NoStop}%
\bibitem [{\citenamefont {Zheludev}\ and\ \citenamefont {Plum}(2016)}]{ref5}%
  \BibitemOpen
  \bibfield  {author} {\bibinfo {author} {\bibfnamefont {N.~I.}\ \bibnamefont
  {Zheludev}}\ and\ \bibinfo {author} {\bibfnamefont {E.}~\bibnamefont
  {Plum}},\ }\href@noop {} {\bibfield  {journal} {\bibinfo  {journal} {Nat.
  Nanotechnol.}\ }\textbf {\bibinfo {volume} {11}},\ \bibinfo {pages} {16}
  (\bibinfo {year} {2016})}\BibitemShut {NoStop}%
\bibitem [{\citenamefont {Tao}\ \emph {et~al.}(2009)\citenamefont {Tao},
  \citenamefont {Strikwerda}, \citenamefont {Fan}, \citenamefont {Padilla},
  \citenamefont {Zhang},\ and\ \citenamefont {Averitt}}]{ref6}%
  \BibitemOpen
  \bibfield  {author} {\bibinfo {author} {\bibfnamefont {H.}~\bibnamefont
  {Tao}}, \bibinfo {author} {\bibfnamefont {A.~C.}~\bibnamefont {Strikwerda}},
  \bibinfo {author} {\bibfnamefont {K.}~\bibnamefont {Fan}}, \bibinfo {author}
  {\bibfnamefont {W.~J.}~\bibnamefont {Padilla}}, \bibinfo {author} {\bibfnamefont
  {X.}~\bibnamefont {Zhang}}, \ and\ \bibinfo {author} {\bibfnamefont
  {R.~D.}~\bibnamefont {Averitt}},\ }\href@noop {} {\bibfield  {journal} {\bibinfo
   {journal} {Phys. Rev. Lett.}\ }\textbf {\bibinfo {volume} {103}},\ \bibinfo
  {pages} {147401} (\bibinfo {year} {2009})}\BibitemShut {NoStop}%
\bibitem [{\citenamefont {Ou}\ \emph {et~al.}(2011)\citenamefont {Ou},
  \citenamefont {Plum}, \citenamefont {Jiang},\ and\ \citenamefont
  {Zheludev}}]{ref7}%
  \BibitemOpen
  \bibfield  {author} {\bibinfo {author} {\bibfnamefont {J.-Y.}\ \bibnamefont
  {Ou}}, \bibinfo {author} {\bibfnamefont {E.}~\bibnamefont {Plum}}, \bibinfo
  {author} {\bibfnamefont {L.}~\bibnamefont {Jiang}}, \ and\ \bibinfo {author}
  {\bibfnamefont {N.~I.}\ \bibnamefont {Zheludev}},\ }\href@noop {} {\bibfield
  {journal} {\bibinfo  {journal} {Nano Lett.}\ }\textbf {\bibinfo {volume}
  {11}},\ \bibinfo {pages} {2142} (\bibinfo {year} {2011})}\BibitemShut
  {NoStop}%
\bibitem [{\citenamefont {Ou}\ \emph {et~al.}(2013)\citenamefont {Ou},
  \citenamefont {Plum}, \citenamefont {Zhang},\ and\ \citenamefont
  {Zheludev}}]{ref8}%
  \BibitemOpen
  \bibfield  {author} {\bibinfo {author} {\bibfnamefont {J.-Y.}\ \bibnamefont
  {Ou}}, \bibinfo {author} {\bibfnamefont {E.}~\bibnamefont {Plum}}, \bibinfo
  {author} {\bibfnamefont {J.}~\bibnamefont {Zhang}}, \ and\ \bibinfo {author}
  {\bibfnamefont {N.~I.}\ \bibnamefont {Zheludev}},\ }\href@noop {} {\bibfield
  {journal} {\bibinfo  {journal} {Nat. Nanotechnol.}\ }\textbf {\bibinfo
  {volume} {8}},\ \bibinfo {pages} {nnano} (\bibinfo {year}
  {2013})}\BibitemShut {NoStop}%
\bibitem [{\citenamefont {Fu}\ \emph {et~al.}(2011)\citenamefont {Fu},
  \citenamefont {Liu}, \citenamefont {Zhu}, \citenamefont {Zhang},
  \citenamefont {Tsai}, \citenamefont {Zhang}, \citenamefont {Mei},
  \citenamefont {Tao}, \citenamefont {Guo}, \citenamefont {Zhang} \emph
  {et~al.}}]{ref9}%
  \BibitemOpen
  \bibfield  {author} {\bibinfo {author} {\bibfnamefont {Y.~H.}\ \bibnamefont
  {Fu}}, \bibinfo {author} {\bibfnamefont {A.~Q.}\ \bibnamefont {Liu}},
  \bibinfo {author} {\bibfnamefont {W.~M.}\ \bibnamefont {Zhu}}, \bibinfo
  {author} {\bibfnamefont {X.~M.}\ \bibnamefont {Zhang}}, \bibinfo {author}
  {\bibfnamefont {D.~P.}\ \bibnamefont {Tsai}}, \bibinfo {author}
  {\bibfnamefont {J.~B.}\ \bibnamefont {Zhang}}, \bibinfo {author}
  {\bibfnamefont {T.}~\bibnamefont {Mei}}, \bibinfo {author} {\bibfnamefont
  {J.~F.}\ \bibnamefont {Tao}}, \bibinfo {author} {\bibfnamefont {H.~C.}\
  \bibnamefont {Guo}}, \bibinfo {author} {\bibfnamefont {X.~H.}\ \bibnamefont
  {Zhang}},  \emph {et~al.},\ }\href@noop {} {\bibfield  {journal} {\bibinfo
  {journal} {Adv. Funct. Mater.}\ }\textbf {\bibinfo {volume} {21}},\ \bibinfo
  {pages} {3589} (\bibinfo {year} {2011})}\BibitemShut {NoStop}%
\bibitem [{\citenamefont {Zhu}\ \emph {et~al.}(2012)\citenamefont {Zhu},
  \citenamefont {Liu}, \citenamefont {Bourouina}, \citenamefont {Tsai},
  \citenamefont {Teng}, \citenamefont {Zhang}, \citenamefont {Lo},
  \citenamefont {Kwong},\ and\ \citenamefont {Zheludev}}]{ref10}%
  \BibitemOpen
  \bibfield  {author} {\bibinfo {author} {\bibfnamefont {W.}~\bibnamefont
  {Zhu}}, \bibinfo {author} {\bibfnamefont {A.}~\bibnamefont {Liu}}, \bibinfo
  {author} {\bibfnamefont {T.}~\bibnamefont {Bourouina}}, \bibinfo {author}
  {\bibfnamefont {D.}~\bibnamefont {Tsai}}, \bibinfo {author} {\bibfnamefont
  {J.}~\bibnamefont {Teng}}, \bibinfo {author} {\bibfnamefont {X.}~\bibnamefont
  {Zhang}}, \bibinfo {author} {\bibfnamefont {G.}~\bibnamefont {Lo}}, \bibinfo
  {author} {\bibfnamefont {D.}~\bibnamefont {Kwong}}, \ and\ \bibinfo {author}
  {\bibfnamefont {N.}~\bibnamefont {Zheludev}},\ }\href@noop {} {\bibfield
  {journal} {\bibinfo  {journal} {Nat. Commun.}\ }\textbf {\bibinfo {volume}
  {3}},\ \bibinfo {pages} {1274} (\bibinfo {year} {2012})}\BibitemShut
  {NoStop}%
\bibitem [{\citenamefont {Valente}\ \emph {et~al.}(2015)\citenamefont
  {Valente}, \citenamefont {Ou}, \citenamefont {Plum}, \citenamefont {Youngs},\
  and\ \citenamefont {Zheludev}}]{ref11}%
  \BibitemOpen
  \bibfield  {author} {\bibinfo {author} {\bibfnamefont {J.}~\bibnamefont
  {Valente}}, \bibinfo {author} {\bibfnamefont {J.-Y.}\ \bibnamefont {Ou}},
  \bibinfo {author} {\bibfnamefont {E.}~\bibnamefont {Plum}}, \bibinfo {author}
  {\bibfnamefont {I.~J.}\ \bibnamefont {Youngs}}, \ and\ \bibinfo {author}
  {\bibfnamefont {N.~I.}\ \bibnamefont {Zheludev}},\ }\href@noop {} {\bibfield
  {journal} {\bibinfo  {journal} {Appl. Phys. Lett.}\ }\textbf {\bibinfo
  {volume} {106}},\ \bibinfo {pages} {111905} (\bibinfo {year}
  {2015})}\BibitemShut {NoStop}%
\bibitem [{\citenamefont {Li}\ \emph {et~al.}(2013)\citenamefont {Li},
  \citenamefont {Shah}, \citenamefont {Withayachumnankul}, \citenamefont {Ung},
  \citenamefont {Mitchell}, \citenamefont {Sriram}, \citenamefont {Bhaskaran},
  \citenamefont {Chang},\ and\ \citenamefont {Abbott}}]{ref12}%
  \BibitemOpen
  \bibfield  {author} {\bibinfo {author} {\bibfnamefont {J.}~\bibnamefont
  {Li}}, \bibinfo {author} {\bibfnamefont {C.~M.}\ \bibnamefont {Shah}},
  \bibinfo {author} {\bibfnamefont {W.}~\bibnamefont {Withayachumnankul}},
  \bibinfo {author} {\bibfnamefont {B.~S.-Y.}\ \bibnamefont {Ung}}, \bibinfo
  {author} {\bibfnamefont {A.}~\bibnamefont {Mitchell}}, \bibinfo {author}
  {\bibfnamefont {S.}~\bibnamefont {Sriram}}, \bibinfo {author} {\bibfnamefont
  {M.}~\bibnamefont {Bhaskaran}}, \bibinfo {author} {\bibfnamefont
  {S.}~\bibnamefont {Chang}}, \ and\ \bibinfo {author} {\bibfnamefont
  {D.}~\bibnamefont {Abbott}},\ }\href@noop {} {\bibfield  {journal} {\bibinfo
  {journal} {Appl. Phys. Lett.}\ }\textbf {\bibinfo {volume} {102}},\ \bibinfo
  {pages} {121101} (\bibinfo {year} {2013})}\BibitemShut {NoStop}%
\bibitem [{\citenamefont {Zhao}\ \emph {et~al.}(2010)\citenamefont {Zhao},
  \citenamefont {Tassin}, \citenamefont {Koschny},\ and\ \citenamefont
  {Soukoulis}}]{ref13}%
  \BibitemOpen
  \bibfield  {author} {\bibinfo {author} {\bibfnamefont {R.}~\bibnamefont
  {Zhao}}, \bibinfo {author} {\bibfnamefont {P.}~\bibnamefont {Tassin}},
  \bibinfo {author} {\bibfnamefont {T.}~\bibnamefont {Koschny}}, \ and\
  \bibinfo {author} {\bibfnamefont {C.~M.}\ \bibnamefont {Soukoulis}},\
  }\href@noop {} {\bibfield  {journal} {\bibinfo  {journal} {Opt. Express}\
  }\textbf {\bibinfo {volume} {18}},\ \bibinfo {pages} {25665} (\bibinfo {year}
  {2010})}\BibitemShut {NoStop}%
\bibitem [{\citenamefont {Volpe}\ \emph {et~al.}(2006)\citenamefont {Volpe},
  \citenamefont {Quidant}, \citenamefont {Badenes},\ and\ \citenamefont
  {Petrov}}]{ref14}%
  \BibitemOpen
  \bibfield  {author} {\bibinfo {author} {\bibfnamefont {G.}~\bibnamefont
  {Volpe}}, \bibinfo {author} {\bibfnamefont {R.}~\bibnamefont {Quidant}},
  \bibinfo {author} {\bibfnamefont {G.}~\bibnamefont {Badenes}}, \ and\
  \bibinfo {author} {\bibfnamefont {D.}~\bibnamefont {Petrov}},\ }\href@noop {}
  {\bibfield  {journal} {\bibinfo  {journal} {Phys. Rev. Lett.}\ }\textbf
  {\bibinfo {volume} {96}},\ \bibinfo {pages} {238101} (\bibinfo {year}
  {2006})}\BibitemShut {NoStop}%
\bibitem [{\citenamefont {Wang}\ \emph {et~al.}(2011)\citenamefont {Wang},
  \citenamefont {Ng}, \citenamefont {Liu}, \citenamefont {Zheng}, \citenamefont
  {Hang},\ and\ \citenamefont {Chan}}]{ref15}%
  \BibitemOpen
  \bibfield  {author} {\bibinfo {author} {\bibfnamefont {S.~B.}~\bibnamefont
  {Wang}}, \bibinfo {author} {\bibfnamefont {J.}~\bibnamefont {Ng}}, \bibinfo
  {author} {\bibfnamefont {H.}~\bibnamefont {Liu}}, \bibinfo {author}
  {\bibfnamefont {H.~H.}~\bibnamefont {Zheng}}, \bibinfo {author} {\bibfnamefont
  {Z.~H.}~\bibnamefont {Hang}}, \ and\ \bibinfo {author} {\bibfnamefont
  {C.~T.}~\bibnamefont {Chan}},\ }\href@noop {} {\bibfield  {journal} {\bibinfo
  {journal} {Phys. Rev. B}\ }\textbf {\bibinfo {volume} {84}},\ \bibinfo
  {pages} {075114} (\bibinfo {year} {2011})}\BibitemShut {NoStop}%
\bibitem [{\citenamefont {Liu}\ \emph {et~al.}(2011)\citenamefont {Liu},
  \citenamefont {Ng}, \citenamefont {Wang}, \citenamefont {Lin}, \citenamefont
  {Hang}, \citenamefont {Chan},\ and\ \citenamefont {Zhu}}]{ref16}%
  \BibitemOpen
  \bibfield  {author} {\bibinfo {author} {\bibfnamefont {H.}~\bibnamefont
  {Liu}}, \bibinfo {author} {\bibfnamefont {J.}~\bibnamefont {Ng}}, \bibinfo
  {author} {\bibfnamefont {S.~B.}~\bibnamefont {Wang}}, \bibinfo {author}
  {\bibfnamefont {Z.~F.}~\bibnamefont {Lin}}, \bibinfo {author} {\bibfnamefont
  {Z.~H.}~\bibnamefont {Hang}}, \bibinfo {author} {\bibfnamefont {C.~T.}~\bibnamefont
  {Chan}}, \ and\ \bibinfo {author} {\bibfnamefont {S.~N.}~\bibnamefont {Zhu}},\
  }\href@noop {} {\bibfield  {journal} {\bibinfo  {journal} {Phys. Rev. Lett.}\
  }\textbf {\bibinfo {volume} {106}},\ \bibinfo {pages} {087401} (\bibinfo
  {year} {2011})}\BibitemShut {NoStop}%
\bibitem [{\citenamefont {Kohoutek}\ \emph {et~al.}(2011)\citenamefont
  {Kohoutek}, \citenamefont {Dey}, \citenamefont {Bonakdar}, \citenamefont
  {Gelfand}, \citenamefont {Sklar}, \citenamefont {Memis},\ and\ \citenamefont
  {Mohseni}}]{ref17}%
  \BibitemOpen
  \bibfield  {author} {\bibinfo {author} {\bibfnamefont {J.}~\bibnamefont
  {Kohoutek}}, \bibinfo {author} {\bibfnamefont {D.}~\bibnamefont {Dey}},
  \bibinfo {author} {\bibfnamefont {A.}~\bibnamefont {Bonakdar}}, \bibinfo
  {author} {\bibfnamefont {R.}~\bibnamefont {Gelfand}}, \bibinfo {author}
  {\bibfnamefont {A.}~\bibnamefont {Sklar}}, \bibinfo {author} {\bibfnamefont
  {O.~G.}\ \bibnamefont {Memis}}, \ and\ \bibinfo {author} {\bibfnamefont
  {H.}~\bibnamefont {Mohseni}},\ }\href@noop {} {\bibfield  {journal} {\bibinfo
   {journal} {Nano Lett.}\ }\textbf {\bibinfo {volume} {11}},\ \bibinfo {pages}
  {3378} (\bibinfo {year} {2011})}\BibitemShut {NoStop}%
\bibitem [{\citenamefont {Lapine}\ \emph {et~al.}(2012)\citenamefont {Lapine},
  \citenamefont {Shadrivov}, \citenamefont {Powell},\ and\ \citenamefont
  {Kivshar}}]{ref18}%
  \BibitemOpen
  \bibfield  {author} {\bibinfo {author} {\bibfnamefont {M.}~\bibnamefont
  {Lapine}}, \bibinfo {author} {\bibfnamefont {I.~V.}\ \bibnamefont
  {Shadrivov}}, \bibinfo {author} {\bibfnamefont {D.~A.}\ \bibnamefont
  {Powell}}, \ and\ \bibinfo {author} {\bibfnamefont {Y.~S.}\ \bibnamefont
  {Kivshar}},\ }\href@noop {} {\bibfield  {journal} {\bibinfo  {journal} {Nat.
  Mater.}\ }\textbf {\bibinfo {volume} {11}},\ \bibinfo {pages} {30} (\bibinfo
  {year} {2012})}\BibitemShut {NoStop}%
\bibitem [{\citenamefont {He}\ \emph {et~al.}(2012)\citenamefont {He},
  \citenamefont {He}, \citenamefont {Gao},\ and\ \citenamefont {Yang}}]{ref19}%
  \BibitemOpen
  \bibfield  {author} {\bibinfo {author} {\bibfnamefont {Y.}~\bibnamefont
  {He}}, \bibinfo {author} {\bibfnamefont {S.}~\bibnamefont {He}}, \bibinfo
  {author} {\bibfnamefont {J.}~\bibnamefont {Gao}}, \ and\ \bibinfo {author}
  {\bibfnamefont {X.}~\bibnamefont {Yang}},\ }\href@noop {} {\bibfield
  {journal} {\bibinfo  {journal} {Opt. Express}\ }\textbf {\bibinfo {volume}
  {20}},\ \bibinfo {pages} {22372} (\bibinfo {year} {2012})}\BibitemShut
  {NoStop}%
\bibitem [{\citenamefont {Zhang}\ \emph {et~al.}(2012)\citenamefont {Zhang},
  \citenamefont {MacDonald},\ and\ \citenamefont {Zheludev}}]{ref20}%
  \BibitemOpen
  \bibfield  {author} {\bibinfo {author} {\bibfnamefont {J.}~\bibnamefont
  {Zhang}}, \bibinfo {author} {\bibfnamefont {K.~F.}~\bibnamefont {MacDonald}}, \
  and\ \bibinfo {author} {\bibfnamefont {N.~I.}~\bibnamefont {Zheludev}},\
  }\href@noop {} {\bibfield  {journal} {\bibinfo  {journal} {Phys. Rev. B}\
  }\textbf {\bibinfo {volume} {85}},\ \bibinfo {pages} {205123} (\bibinfo
  {year} {2012})}\BibitemShut {NoStop}%
\bibitem [{\citenamefont {Ginis}\ \emph {et~al.}(2013)\citenamefont {Ginis},
  \citenamefont {Tassin}, \citenamefont {Soukoulis},\ and\ \citenamefont
  {Veretennicoff}}]{ref21}%
  \BibitemOpen
  \bibfield  {author} {\bibinfo {author} {\bibfnamefont {V.}~\bibnamefont
  {Ginis}}, \bibinfo {author} {\bibfnamefont {P.}~\bibnamefont {Tassin}},
  \bibinfo {author} {\bibfnamefont {C.~M.}\ \bibnamefont {Soukoulis}}, \ and\
  \bibinfo {author} {\bibfnamefont {I.}~\bibnamefont {Veretennicoff}},\
  }\href@noop {} {\bibfield  {journal} {\bibinfo  {journal} {Phys. Rev. Lett.}\
  }\textbf {\bibinfo {volume} {110}},\ \bibinfo {pages} {057401} (\bibinfo
  {year} {2013})}\BibitemShut {NoStop}%
\bibitem [{\citenamefont {Marcet}\ \emph {et~al.}(2014)\citenamefont {Marcet},
  \citenamefont {Hang}, \citenamefont {Wang}, \citenamefont {Ng}, \citenamefont
  {Chan},\ and\ \citenamefont {Chan}}]{ref22}%
  \BibitemOpen
  \bibfield  {author} {\bibinfo {author} {\bibfnamefont {Z.}~\bibnamefont
  {Marcet}}, \bibinfo {author} {\bibfnamefont {Z.~H.}~\bibnamefont {Hang}},
  \bibinfo {author} {\bibfnamefont {S.~B.}~\bibnamefont {Wang}}, \bibinfo {author}
  {\bibfnamefont {J.}~\bibnamefont {Ng}}, \bibinfo {author} {\bibfnamefont
  {C.~T.}~\bibnamefont {Chan}}, \ and\ \bibinfo {author} {\bibfnamefont
  {H.~B.}~\bibnamefont {Chan}},\ }\href@noop {} {\bibfield  {journal} {\bibinfo
  {journal} {Phys. Rev. Lett.}\ }\textbf {\bibinfo {volume} {112}},\ \bibinfo
  {pages} {045504} (\bibinfo {year} {2014})}\BibitemShut {NoStop}%
\bibitem [{\citenamefont {Ma}\ \emph {et~al.}(2015)\citenamefont {Ma},
  \citenamefont {Garrett},\ and\ \citenamefont {Munday}}]{ref23}%
  \BibitemOpen
  \bibfield  {author} {\bibinfo {author} {\bibfnamefont {D.}~\bibnamefont
  {Ma}}, \bibinfo {author} {\bibfnamefont {J.~L.}\ \bibnamefont {Garrett}}, \
  and\ \bibinfo {author} {\bibfnamefont {J.~N.}\ \bibnamefont {Munday}},\
  }\href@noop {} {\bibfield  {journal} {\bibinfo  {journal} {Appl. Phys.
  Lett.}\ }\textbf {\bibinfo {volume} {106}},\ \bibinfo {pages} {091107}
  (\bibinfo {year} {2015})}\BibitemShut {NoStop}%
\bibitem [{\citenamefont {Guan}\ \emph {et~al.}(2015)\citenamefont {Guan},
  \citenamefont {Hang}, \citenamefont {Marcet}, \citenamefont {Liu},
  \citenamefont {Kravchenko}, \citenamefont {Chan}, \citenamefont {Chan},\ and\
  \citenamefont {Tong}}]{ref24}%
  \BibitemOpen
  \bibfield  {author} {\bibinfo {author} {\bibfnamefont {D.}~\bibnamefont
  {Guan}}, \bibinfo {author} {\bibfnamefont {Z.~H.}\ \bibnamefont {Hang}},
  \bibinfo {author} {\bibfnamefont {Z.}~\bibnamefont {Marcet}}, \bibinfo
  {author} {\bibfnamefont {H.}~\bibnamefont {Liu}}, \bibinfo {author}
  {\bibfnamefont {I.~I.}\ \bibnamefont {Kravchenko}}, \bibinfo {author}
  {\bibfnamefont {C.~T.}\ \bibnamefont {Chan}}, \bibinfo {author}
  {\bibfnamefont {H.~B.}\ \bibnamefont {Chan}}, \ and\ \bibinfo {author}
  {\bibfnamefont {P.}~\bibnamefont {Tong}},\ }\href@noop {} {\bibfield
  {journal} {\bibinfo  {journal} {Sci. Rep.}\ }\textbf {\bibinfo {volume}
  {5}},\ \bibinfo {pages} {16216} (\bibinfo {year} {2015})}\BibitemShut
  {NoStop}%
\bibitem [{\citenamefont {Ou}\ \emph {et~al.}(2016)\citenamefont {Ou},
  \citenamefont {Plum}, \citenamefont {Zhang},\ and\ \citenamefont
  {Zheludev}}]{ref25}%
  \BibitemOpen
  \bibfield  {author} {\bibinfo {author} {\bibfnamefont {J.-Y.}\ \bibnamefont
  {Ou}}, \bibinfo {author} {\bibfnamefont {E.}~\bibnamefont {Plum}}, \bibinfo
  {author} {\bibfnamefont {J.}~\bibnamefont {Zhang}}, \ and\ \bibinfo {author}
  {\bibfnamefont {N.~I.}\ \bibnamefont {Zheludev}},\ }\href@noop {} {\bibfield
  {journal} {\bibinfo  {journal} {Adv. Mater.}\ }\textbf {\bibinfo {volume}
  {28}},\ \bibinfo {pages} {729} (\bibinfo {year} {2016})}\BibitemShut
  {NoStop}%
\bibitem [{\citenamefont {Jin}\ \emph {et~al.}(2016)\citenamefont {Jin},
  \citenamefont {Li}, \citenamefont {Wang}, \citenamefont {Zhu}, \citenamefont
  {Li},\ and\ \citenamefont {Dong}}]{ref26}%
  \BibitemOpen
  \bibfield  {author} {\bibinfo {author} {\bibfnamefont {R.-c.}\ \bibnamefont
  {Jin}}, \bibinfo {author} {\bibfnamefont {J.}~\bibnamefont {Li}}, \bibinfo
  {author} {\bibfnamefont {Y.-h.}\ \bibnamefont {Wang}}, \bibinfo {author}
  {\bibfnamefont {M.-j.}\ \bibnamefont {Zhu}}, \bibinfo {author} {\bibfnamefont
  {J.-q.}\ \bibnamefont {Li}}, \ and\ \bibinfo {author} {\bibfnamefont {Z.-g.}\
  \bibnamefont {Dong}},\ }\href@noop {} {\bibfield  {journal} {\bibinfo
  {journal} {Opt. Express}\ }\textbf {\bibinfo {volume} {24}},\ \bibinfo
  {pages} {27563} (\bibinfo {year} {2016})}\BibitemShut {NoStop}%
\bibitem [{\citenamefont {Zhang}\ \emph {et~al.}(2014)\citenamefont {Zhang},
  \citenamefont {MacDonald},\ and\ \citenamefont {Zheludev}}]{ref27}%
  \BibitemOpen
  \bibfield  {author} {\bibinfo {author} {\bibfnamefont {J.}~\bibnamefont
  {Zhang}}, \bibinfo {author} {\bibfnamefont {K.~F.}\ \bibnamefont
  {MacDonald}}, \ and\ \bibinfo {author} {\bibfnamefont {N.~I.}\ \bibnamefont
  {Zheludev}},\ }\href@noop {} {\bibfield  {journal} {\bibinfo  {journal} {Opt.
  Lett.}\ }\textbf {\bibinfo {volume} {39}},\ \bibinfo {pages} {4883} (\bibinfo
  {year} {2014})}\BibitemShut {NoStop}%
\bibitem [{\citenamefont {Wang}\ and\ \citenamefont {Chan}(2013)}]{ref28}%
  \BibitemOpen
  \bibfield  {author} {\bibinfo {author} {\bibfnamefont {S.~B.}\ \bibnamefont
  {Wang}}\ and\ \bibinfo {author} {\bibfnamefont {C.~T.}\ \bibnamefont
  {Chan}},\ }\href@noop {} {\bibfield  {journal} {\bibinfo  {journal} {J. Phys.
  D: Appl. Phys.}\ }\textbf {\bibinfo {volume} {46}},\ \bibinfo {pages}
  {395104} (\bibinfo {year} {2013})}\BibitemShut {NoStop}%
\bibitem [{\citenamefont {Sun}\ \emph {et~al.}(2015)\citenamefont {Sun},
  \citenamefont {Wang}, \citenamefont {Ng}, \citenamefont {Zhou},\ and\
  \citenamefont {Chan}}]{ref29}%
  \BibitemOpen
  \bibfield  {author} {\bibinfo {author} {\bibfnamefont {W.}~\bibnamefont
  {Sun}}, \bibinfo {author} {\bibfnamefont {S.~B.}~\bibnamefont {Wang}}, \bibinfo
  {author} {\bibfnamefont {J.}~\bibnamefont {Ng}}, \bibinfo {author}
  {\bibfnamefont {L.}~\bibnamefont {Zhou}}, \ and\ \bibinfo {author}
  {\bibfnamefont {C.~T.}\ \bibnamefont {Chan}},\ }\href@noop {} {\bibfield
  {journal} {\bibinfo  {journal} {Phys. Rev. B}\ }\textbf {\bibinfo {volume}
  {91}},\ \bibinfo {pages} {235439} (\bibinfo {year} {2015})}\BibitemShut
  {NoStop}%
\bibitem [{\citenamefont {Wang}\ \emph {et~al.}(2016)\citenamefont {Wang},
  \citenamefont {Ng}, \citenamefont {Xiao},\ and\ \citenamefont
  {Chan}}]{ref30}%
  \BibitemOpen
  \bibfield  {author} {\bibinfo {author} {\bibfnamefont {S.}~\bibnamefont
  {Wang}}, \bibinfo {author} {\bibfnamefont {J.}~\bibnamefont {Ng}}, \bibinfo
  {author} {\bibfnamefont {M.}~\bibnamefont {Xiao}}, \ and\ \bibinfo {author}
  {\bibfnamefont {C.~T.}\ \bibnamefont {Chan}},\ }\href@noop {} {\bibfield
  {journal} {\bibinfo  {journal} {Sci. Adv.}\ }\textbf {\bibinfo {volume}
  {2}},\ \bibinfo {pages} {e1501485} (\bibinfo {year} {2016})}\BibitemShut
  {NoStop}%
\bibitem [{\citenamefont {Metzger}\ and\ \citenamefont {Karrai}(2004)}]{ref31}%
  \BibitemOpen
  \bibfield  {author} {\bibinfo {author} {\bibfnamefont {C.~H.}\ \bibnamefont
  {Metzger}}\ and\ \bibinfo {author} {\bibfnamefont {K.}~\bibnamefont
  {Karrai}},\ }\href@noop {} {\bibfield  {journal} {\bibinfo  {journal}
  {Nature}\ }\textbf {\bibinfo {volume} {432}},\ \bibinfo {pages} {1002}
  (\bibinfo {year} {2004})}\BibitemShut {NoStop}%
\bibitem [{\citenamefont {Leissa}(1969)}]{ref32}%
  \BibitemOpen
  \bibfield  {author} {\bibinfo {author} {\bibfnamefont {A.~W.}\ \bibnamefont
  {Leissa}},\ }\href@noop {} {\emph {\bibinfo {title} {Vibration of plates}}},\
  \bibinfo {type} {Tech. Rep.}\ (\bibinfo  {institution} {OHIO STATE UNIV
  COLUMBUS},\ \bibinfo {year} {1969})\BibitemShut {NoStop}%
\end{thebibliography}

\end{document}